\documentclass[letterpaper]{jpconf}
\usepackage{graphicx}
\usepackage{cite}

\begin{document}
\title{Direct Determination of  Neutrino Mass}

\author{R.G. Hamish Robertson}

\address{Center for Experimental Nuclear Physics and Astrophysics, University of Washington, Seattle, WA, 98195, USA}

\ead{rghr@u.washington.edu}

%\author{KATRIN Collaboration}

\begin{abstract}
The sum of the masses of the three neutrino mass eigenstates is now constrained both from above and below, and lies between 55 and 6900 meV.  The lower limit is set by neutrino oscillations and the fact that masses are non-negative.  The upper limit is set by laboratory measurements on the beta decay of tritium.  These determinations  share a common characteristic of being essentially model independent, or ``direct''.   The clustering on large scales in the universe depends on this quantity, and, within the framework of $\Lambda$CDM cosmology, favors a mass sum below about 600 meV.  In this article, the laboratory approach to neutrino mass via beta decay is emphasized, particularly an experiment now under construction, KATRIN, on the beta decay of tritium.   Another candidate beta-active nuclide, $^{187}$Re, offers an advantage in phase space but calls for a very different experimental approach.

\end{abstract}

\section{Introduction}

While it is now certain from neutrino oscillation experiments that there are three neutrinos with different masses, the absolute masses  remain unknown.  The masses are of importance to cosmology, because neutrino mass influences the formation of large-scale structure in the universe, tending to damp clustering until an epoch when the neutrinos have cooled sufficiently to become non-relativistic.   The scale-dependence of clustering observed in the universe, interpreted in the framework of a $\Lambda$CDM cosmology, favors a mass sum below about 600 meV \cite{Fogli:2008ig}.  A laboratory determination of the mass would be valuable in constraining cosmological models, because there are significant correlations between neutrino mass and both the running of the spectral index and the equation of state of dark energy \cite{Fogli:2006yq}.

Neutrino mass is also an important issue in the development of the next standard model.  In this case, it is the {\em patterns} of mass rather than the numerical details that can be expected to shed light on the mechanisms of mass generation.  A quasi-degenerate, an inverted hierarchical, and a normal hierarchical spectrum would each send a unique message about the role of mass and possible new symmetries beyond the standard model.

\section{Oscillations}

\begin{figure}
\begin{center}
\includegraphics[width=5in]{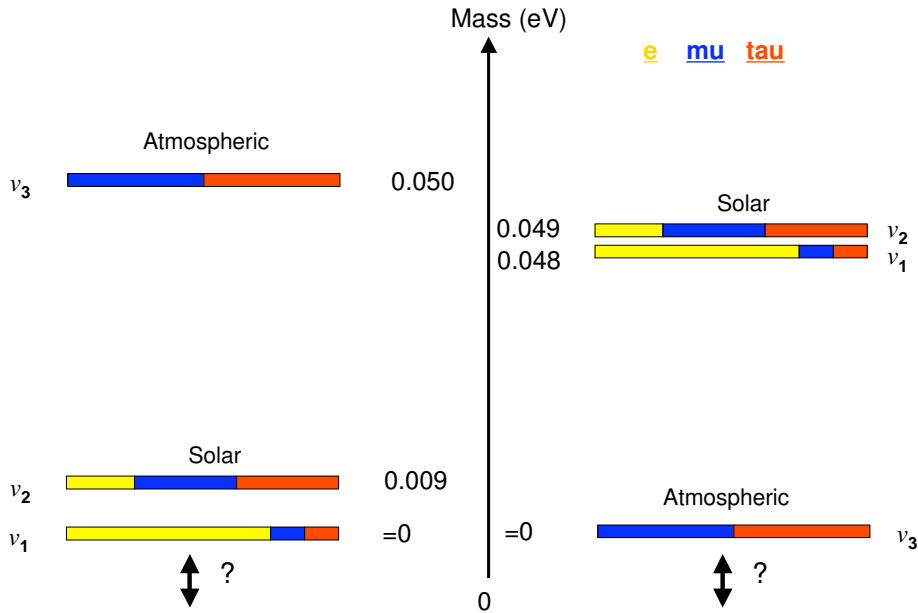}
\end{center}
\vspace*{.5\baselineskip}
\caption{
Schematic illustration (mass intervals not to scale, mass values given for lightest mass = 0) of the decomposition of the neutrino mass eigenstates $\nu_i$ in terms of their flavor intensities. There may also be a small admixture of $\nu_e$ into $\nu_3$.  Which of the two hierarchies is correct is not at present known. }
\label{fig:massspectrum1}
\end{figure}

Neutrino oscillations yield the differences between the squares of the eigenmasses.  Two such differences (`solar' and `atmospheric') have been measured in a variety of experiments on solar, atmospheric, reactor, and accelerator neutrinos.  Given that the differences must close cyclically, this is sufficient to determine all three such quantities, except that the differences are signed quantities.  The sign is known for the solar mass gap $\delta m^2$ from the effects of matter enhancement, but it is not known for the atmospheric gap $\Delta m^2$.  The two possibilities lead to the `normal' and `inverted' hierarchy, as illustrated in Fig.~\ref{fig:massspectrum1}.
The most recent determinations yield \cite{Fogli:2008ig} the following values (95\% CL) for the mass-squared differences: 
\begin{eqnarray}
\delta m^2 & = & (7.66\pm0.35)\times10^{-5}{\rm\ eV}^2 \\
\Delta m^2 & = & (2.38\pm0.27)\times10^{-3}{\rm\ eV}^2 .
\end{eqnarray}
The smallest possible sum (with 1-$\sigma$ uncertainties) of the three masses is found in the normal hierarchy, with $m_1 = 0$: 
\begin{eqnarray}
\sum m_i &  \geq & (58.3^{+1.4}_{-1.5}){\rm\ meV} 
\end{eqnarray}
If the hierarchy is inverted, the smallest possible sum would be,
\begin{eqnarray}
\sum m_i &  \geq & (97.8^{+2.8}_{-2.8}){\rm\ meV} 
\end{eqnarray}
The precision of the lower bound on the sum of the masses is remarkable.

\section{Beta Decay}

In beta decay the electron and neutrino share the available energy in a statistical fashion, and in a small fraction of the decays the electron will take almost all the energy unless some must be reserved for the rest mass of the neutrino.   The phase-space available to the electron near the endpoint is therefore modified by neutrino mass, a fact realized immediately by Fermi \cite{Fermi:1934sk} as he formulated the theory of beta decay.

In the intervening seven decades of experimental searches for neutrino mass, tritium has been the beta-active nucleus of choice because it has a low endpoint energy, which makes the modification caused by neutrino mass a larger fraction of the total spectrum:  
\begin{displaymath}
  ^3{\rm H}\,\rightarrow \, ^3{\rm He}^+\,+\,{\rm e}^- \,+\,\overline{\nu}_{\rm e}  \quad {\rm +\ 18580\  eV.}
\end{displaymath}
It is also a very simple atom, and atomic or molecular effects are important at the eV level.  The decay is superallowed, with a short half-life, which reduces the amount of source material needed for a given counting rate and sensitivity. 

A recent complete treatment of the spectrum shape in beta decay is given by Masood {\em et al.} \cite{Masood:2007rc}.  In the standard analysis, the electron energy spectrum for neutrinos  with eigenmasses $m_i$ is given to a good approximation by
\begin{equation}
{dN \over dT} = CF(Z,T) p (T+m_e) (T_0-T) \sum_{i=1,3}|U_{ei}|^2[(T_0-T)^2-m_i^2]^{1
\over 2} \Theta (T_0-T-m_i), \label{mother}
\end{equation}
where $T$ denotes the kinetic energy of the electron, $U_{ei}$ is the electron-flavor amplitude in  the $i$th mass eigenstate, $m_e$ is the
mass of the electron, $p$ is the electron momentum,  $T_0$
corresponds to the maximum electron energy for $m_i=0$ (endpoint
energy), $F(Z,T)$ is the Fermi function, 
and the step function $\Theta (T_0-T-m_i)$ ensures energy
conservation.
 The constant $C$ is given by 
\begin{equation}  C={G_F^2  \over 2 \pi^3}
\cos^2 \theta_C |M|^2~. \label{re}
\end{equation}
Here $G_F$ is the Fermi
constant, $\theta_C$ is the Cabibbo angle and $M$ is the nuclear
matrix element. 
As both  $M$ and $F(Z,T)$ are independent of
$m_i$, the dependence of the spectral shape on $m_i$ is given
by the phase space factor only. Moreover, the bound on the
neutrino mass from beta decay is independent of whether the
electron neutrino is a Majorana or a Dirac particle.

Above 200 meV the masses $m_i$ are essentially equal and degenerate, the unitarity of {\bf U} gives $\sum_{i}|U_{ei}|^2=1$, and the resulting beta spectrum is indistinguishable from what  a single massive `electron neutrino'  would produce if mass eigenstates were also flavor eigenstates.     The current best limit is that reported by the Mainz group, $m_i \leq 2300$ meV (95\% CL) \cite{Kraus:2004zw}.  The corresponding limit on the sum of the masses is 6900 meV, and the sum is therefore known from model-independent constraints to lie in the range,
$$55.3 \leq \sum m_i \leq 6900 {\rm \ meV}$$
at the 95\% CL.

\section{Future Beta Decay Experiments}

Two beta-active nuclides are attractive for the next phase of experimental work.  Tritium will be the basis for the next-generation KATRIN experiment now under construction in Germany \cite{Angrik:2005ep},  and $^{187}$Re is the basis of the MARE project in Italy \cite{Andreotti:2007eq}.  These nuclides have the lowest endpoint energies.   Table \ref{tab:betaemitters} summarizes some of the basic quantities relevant to each.

Because the spectrum near the endpoint is quadratic in both cases, $^{187}$Re enjoys a factor of about 300 advantage over tritium in the branching ratio to the last eV because of the low endpoint energy of the Re.  On the other hand, the forbidden nature of the decay and the correspondingly low specific activity means that to obtain one event per day in the last 200 meV of the spectrum requires 20 $\mu$g of tritium, but 13 kg of $^{187}$Re.  If nothing is seen in the current round of experiments down to 200 meV, it will be necessary to develop a program with 20 meV sensitivity, for which the corresponding source masses are 20 mg of tritium and 13 tonnes of $^{187}$Re.  

\begin{center}
\begin{table}[ht]
\caption{\label{tab:betaemitters} Parameters of $^3$H and $^{187}$Re (taken to be isotopically pure for all listings except isotopic abundance).  Each nuclide is treated as though having only a single final state.}
%\footnotesize\rm
\centering
\begin{tabular}{@{}*{7}{lccl}}
\br
Parameter & $^3$H	& $^{187}$Re & Unit \\
\mr
Spin sequence & ${1/2}^+\rightarrow {1/2}^+$ & ${5/2}^+\rightarrow {1/2}^-$&\\
Half-life & 12.32 & $4.32\times10^{10}$ & y \\
Isotopic abundance & -- & 62.6 & \% \\
Endpoint energy & 18.58 & 2.47 & keV \\
Specific activity & $3.58\times 10^{14}$ & $1.64\times 10^{3}$  & Bq/g \\
Branch to last eV &  $2\times 10^{-13}$  & $6\times 10^{-11}$ & \\
Specific activity, last eV & 72 & $1.1\times 10^{-7}$  & Bq/g \\
\br
\end{tabular}
\end{table}
\end{center}

The experimental approaches in the two cases differ fundamentally.  For tritium experiments the source and detector are separate, which permits filtering out the bulk of uninteresting decays before they reach the detector, but which also sets a limit on how large the source can be while still getting the electrons out of it. For $^{187}$Re, on the other hand, the experiments are done with microcalorimeters in order to capture all the energy, independent of the complexities of atomic final state excitations.  In that case the source and detector are the same, obviating the fundamental limit for separated source and detector, but raising the problem that the detector must detect every event no matter whether it is interesting or not.  Avoidance of spectral distortions and background caused by pileup then becomes vitally important.   

\section{The MARE Experiment}

MARE, Microcalorimeter Arrays for a Rhenium Experiment, brings together the technical advances of earlier programs, MANU \cite{Pergolesi:2006qw} and MIBETA \cite{Sisti:2004iq,Arnaboldi:2006xxx}, with a view to optimizing the approach for the next phases \cite{Andreotti:2007eq}.  MARE-1 aims for 2-eV sensitivity, and MARE-2 200 meV.  The MANU approach involves superconducting Re crystals and transition-edge sensors made of an Ir-Au alloy, while the MIBETA approach has used crystals of AgReO$_4$ and resistance thermometers made of doped Si.  MIBETA has measured the endpoint energy of the $^{187}$Re beta decay to be 
$$T_0 = (2465.3 \pm 0.5 \pm 1.6) {\rm \ eV},$$ and the the neutrino mass limit to be
$$m_\nu^2= (-112 \pm 207 \pm 90) {\rm \ eV}^2,$$corresponding to an upper limit of 15 eV on the mass.

\section{The KATRIN Experiment\footnote[1]{These sections are an updated version of an earlier presentation on the KATRIN experiment \cite{Robertson:2007xx}.
}}

The KArlsruhe TRItium Neutrino experiment \cite{Angrik:2005ep} is a very large scale tritium-beta-decay experiment to determine the mass of the neutrino.  It is presently under construction at the Forschungszentrum Karlsruhe, and makes use of the Tritium Laboratory built there for the ITER project.  The combination of a very large retarding-potential electrostatic-magnetic spectrometer and an intense gaseous molecular tritium source makes possible a sensitivity to neutrino mass of 0.2 eV, about an order of magnitude below present laboratory limits.

\begin{figure}[hb]
\begin{center}
\includegraphics[width=6in]{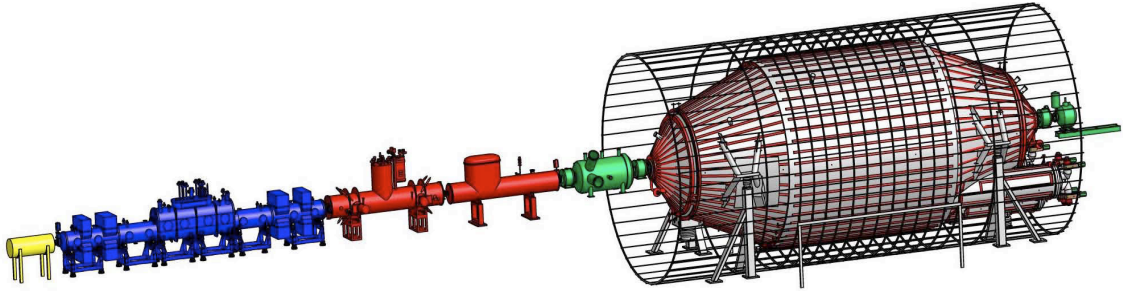}
\end{center}
\caption{\label{fig:overview}Layout of the KATRIN Experiment}
\end{figure}

KATRIN consists of seven major subsystems (see Fig.~\ref{fig:overview}), a gaseous tritium source, the tritium processing and recirculation system, a differential pumping section, a cryogenic Ar frost pumping section, a pre-spectrometer, a main spectrometer, and the detector system.  

\section{Source and Tritium Recirculation}
The gaseous molecular source consists of a tube 9 cm in diameter and 10 m in length maintained at the desired temperature (27 K) by circulation of two-phase Ne.  An axial magnetic field of 3.6 T guides electrons toward the spectrometers.   Four turbomolecular pumps at the ends of the source tube collect T$_2$ for recirculation in an `inner loop' with Pd filters to remove contaminants, particularly $^3$He.  A `demonstrator' for the source is being constructed by ACCEL Instruments GmbH to verify the thermal performance. 

\section{Differential Pumping Section}
Further reduction of tritium pressure is achieved in this section, which consists of a tube with a chicane, superconducting solenoids to guide the electrons, and four 2800 l/s magnetic-bearing turbomolecular pumps to scavenge the tritium.   The subsystem is being built by Ansaldo Superconduttori for delivery in 2008.
\section{Cryogenic Pumping Section}
Any remaining tritium that escapes through the differential pumping section is trapped in Ar frost, which forms a highly efficient, large-area, chemically inert, and radiation-immune surface.  The contract for this system is in place.
\section{Pre-spectrometer}
There are two spectrometers in tandem in KATRIN,  both of the retarding-potential type. The pre-spectrometer operates at a cutoff potential typically 100 eV below the endpoint, preventing most electrons from reaching the main spectrometer.  Electrons can ionize residual gas molecules and create slow electrons that are indistinguishable from the signal. Reducing the electron flux into the main spectrometer is expected to improve the background near the endpoint substantially.  The pre-spectrometer was the first KATRIN subassembly to be completed, in order to permit some key concepts to be tested before other design elements were frozen.  One important discovery was the existence of a parasitic Penning-trap configuration near the ends, which made it impossible to apply high voltage and a magnetic field simultaneously without breakdown.  The addition of appropriate  electrodes suppressed this discharge completely, and the spectrometer now runs uneventfully at  35 kV and 4 T. A new feature in this type of spectrometer is a grid inside the shell to suppress the emission of low-energy electrons ejected by cosmic rays and radioactivity from the shell.  The performance of the grid  will be tested shortly, to verify the design principles of the much larger grid system for the main spectrometer.  
\section{Main Spectrometer}
The main spectrometer is a large stainless-steel vessel 10 m in diameter and 24 m in length.  The 200-tonne chamber was fabricated in Deggendorf by MAN-DWE and shipped via the river Danube, the Black Sea, the Mediterranean, the North Sea, and the Rhine to be offloaded at Leopoldshafen near the FZK laboratory.  It was placed in its building November 29,  2006.  Thermal insulation and heating tubes were installed, and 6 turbopumps. The tank was baked out to 350 C from July 16, 2007 to July 25.  The base pressure 2 months later was $0.9\times10^{-9}$ mbar and the outgassing rate (predominantly H$_2$) $1.2\times10^{-12}$ mbar l s$^{-1}$ cm$^{-2}$.  When the interior grid structure and 1000 m of non-evaporable getter strips are installed later, the predicted base pressure will be $3\times10^{-11}$ mbar.  
\section{Detector}
Electrons surmounting the potential barriers in the spectrometers enter a high axial magnetic field region and are detected in a monolithic 148-pixel Si PIN diode array 10 cm in diameter.  Two superconducting solenoids can produce up to 6 T, defining the electron beam diameter and the maximum pitch angle accepted from the source.  The magnets are being made by Cryomagnetics, Inc.~and the detector by Canberra. Between the solenoids is a region in which various calibration devices -- gamma sources and a photoemissive electron gun -- can be inserted.  Selection of materials, shielding,  and an active veto are being adopted to keep non-beam-associated backgrounds below 1 mHz.  
\section{Summary}
The KATRIN experiment is scheduled for initial data-taking in 2010.  Subsystems are in an advanced state of construction and commissioning.  Funding for KATRIN is being provided by the Helmholtz Gemeinschaft, the Bundesministerium f\"{u}r Bildung und Forschung, and the US Department of Energy.  MARE is in the R\&D phase, with support from INFN, Italy.

This paper was prepared for the Carolina International Symposium on Neutrino Physics, Columbia, SC, May 15 - 17, 2008, and the author  acknowledges with great respect  the advances in neutrino physics springing from the work of Frank T. Avignone, Ettore Fiorini, and the late S. Peter Rosen.  This research was supported by the US DOE under Grant DE-FG02-97ER41020.

\section*{References}
\bibliographystyle{iopart-num}
\bibliography{testbib}{}

%\begin{thebibliography}{9}
%\bibitem{iopartnum} IOP Publishing is to grateful Mark A Caprio, Center for Theoretical Physics, Yale University, for permission to include the {\tt iopart-num} \BibTeX package (version 2.0, December 21, 2006) with  this documentation. Updates and new releases of {\tt iopart-num} can be found on \verb"www.ctan.org" (CTAN). 
%\end{thebibliography}

\end{document}